\documentclass[12pt,preprint]{aastex}
\shorttitle{PCA of Galactic Globular Clusters}
\shortauthors{Strader \& Brodie}
\def\etal{{\it et al.}}
\input epsf

\begin{document}

\title{A Principal Components Analysis of the Lick Indices of Galactic Globular Clusters}

\author{Jay Strader and Jean P. Brodie}
\affil{UCO/Lick Observatory, University of California, Santa Cruz, CA 95064}
\email{strader@ucolick.org, brodie@ucolick.org}

\begin{abstract}

We present a principal components analysis (PCA) of high-quality Lick/IDS absorption-line measurements for 11 indices in the wavelength range 4100--5400 \AA\ for 39 Galactic
globular clusters (GCs). Only the first principal component appears to be physically significant. We find that there is a tight linear relationship between this first component
(PC1) and GC metallicity over a wide range in [m/H] ($-1.8 \le$ [m/H] $\le 0$), suggesting that PC1 may be used to accurately estimate metallicities for old extragalactic GCs
from their integrated spectra.\footnote{Metallicity calibrations for other sets of indices in this wavelength range can be furnished upon request.} The classic GC ``second
parameter effect'' is noticeable in the Balmer indices, though it does not appear in our PCA. We find little evidence for substantial differences in broad abundance patterns
among Galactic GCs. One implication is that the metal-poor and metal-rich GC subpopulations formed from very similar physical processes.
 
\end{abstract}


\section{Introduction}

A decade ago, the discovery of bimodality in the color distributions of globular clusters (GCs) in luminous galaxies (Zepf \& Ashman 1993), coupled with evidence that proto-GCs
were forming in major star formation events such as major mergers (Holtzman \etal~1992; Whitmore \etal~1993), led to the suggestion that most massive galaxies underwent at least
two significant episodes of star formation. In the Milky Way, the two GC subpopulations appear to have very similar ages, implying that the color bimodality is due entirely to
metallicity differences. Increasingly, spectroscopic studies of GCs in external galaxies (primarily Es and S0s) indicate that the bulk of their GCs are also old (e.g., Larsen \&
Brodie~2002), although a small number of intermediate-age GCs may sometimes be present (Goudfrooij \etal~2001; Strader \etal~2003). Thus, in many galaxies GC bimodality appears to
be attributable to metallicity.

These GC subpopulations may hold important insights into galaxy formation. The metallicities of the metal-rich GCs closely track those of the bulge population of their parent
galaxy (McWilliam \& Rich 1994; Forbes, Brodie, \& Grillmair 1997; Forbes, Brodie, \& Larsen 2001). Recent work shows that the mean metallicity of the metal-poor GCs, which was
generally considered to be approximately uniform among galaxies, is also correlated with parent galaxy mass (Larsen \etal~2001; Strader, Brodie, \& Forbes 2004). These findings
underscore the importance of deriving accurate metallicities for extragalactic GCs. One such significant effort was that of Brodie \& Huchra (1990), who estimated GC metallicities
from a set of nine absorption-line indices in the blue.

It has become common to estimate metallicities and ages for extragalactic GCs using Lick ``index-index'' grids (Gonz{\' a}lez 1993), in which a metallicity-sensitive index is
plotted against an age-sensitive index, with simple stellar population (SSP) models overlaid. However, this procedure is far from ideal, as it uses only a small fraction of the
information available in the spectrum. We decided to use principal components analysis (PCA; see \S 2) to (i) explore the simultaneous use of a large number of absorption lines
in estimating metallicity, and (ii) determine whether factors other than metallicity significantly affect the measured Lick indices of Galactic GCs. For example, if the
metal-poor and metal-rich GC subpopulations formed through different physical processes, this might be reflected in distinct chemical abundance patterns.

A PCA of 26 Galactic GCs using nine Lick indices measured with the original Burstein \etal~(1984) definitions was performed by Covino, Galletti, \& Pasinetti (1995). Their main
finding was that the Lick indices of their sample of GCs appeared to be governed by a single physical parameter (metallicity), but they did not calibrate this component for use
as a metallicity estimator. A non-PCA spectroscopic analysis of the central regions of several metal-rich GCs by Rose and Tripicco (1986) found evidence for a second parameter,
which demonstrated itself in a varying dwarf-to-giant ratios (estimated with a \ion{Sr}{2}/\ion{Fe}{1} line ratio) and CN anomalies. They suggested that the most plausible
cause for this second parameter was a variation in the numbers of coalesced binaries in the cores of these GCs. Gregg (1994) presented a more extensive study of the integrated
spectra of 13 Galactic GCs. Although his spectra had modest resolution ($\sim 5$ \AA), he measured many absorption-line indices rarely considered in other studies. We compare
his results to ours in \S 3 and \S 4 below.

\section{The Method and Sample}

PCA is one of the most common multivariate statistical methods used in astronomy (see Francis \& Wills 1999 for a good introduction). A short description follows. Assume one has
measured $N$ parameters for each of a set of objects. PCA constructs a set of $N$ independent, orthogonal variables that are linear combinations of the input parameters, such that
each of these new variables (which are called principal components) in turn accounts for as much of the variance in the entire data set as possible.  This is done algebraically by
finding the eigenvectors and eigenvalues of the covariance matrix. Geometrically, one could imagine the data as a cloud of points in $N$-dimensional space. If these quantities are
uncorrelated, one would expect the points to fill space more or less uniformly, and all of the principal components would account for approximately equal percentages of the
variance. However, if a subset of the parameters are correlated, the cloud of points would be elongated in this space.  In this case, the first principal component (typically
called PC1) would be the least-squares line through the cloud, and would then represent the first axis of the new basis set being constructed. This process is then repeated in the
$N-1$-dimensional space orthogonal to PC1. In actuality, all PCs are determined simultaneously, but the serial procedure described here offers a good conceptual sense of the
process.

It may often be the case that the bulk of the variance in the data can be accounted for by the first few PCs; in this case, a $N$-dimensional problem has been vastly simplified.
Generally, the input parameters will not have been measured in the same units, so it is necessary to normalize them to have the same mean and variance (usually taken to be zero
and one) before performing PCA. It should be noted that the PCs may have no clear physical interpretation, even if highly significant. Also, since PCA is a linear analysis,
nonlinear relationships between the variables cannot be expressed (though they can sometimes be accommodated with an appropriate variable transformation).

We used Lick/IDS index measurements for 39 Galactic GCs from Schiavon \etal~(2004). They measured indices using the passband definitions for 11 of the ``base'' indices given in
Worthey \etal~(1994). Four higher-order Balmer line indices have also been measured using the definitions in Worthey \& Ottaviani (1997). From this total set of 15 indices, we
excluded Fe5015, because of large scatter which appears to be intrinsic. We also excluded CN$_1$, since it measures the same features as CN$_2$ but can suffer from H$\delta$
contamination. The wider ``A'' passband definitions of H$\gamma$ and H$\delta$ were used in favor of the two ``F'' indices. If the ``F'' indices are used instead, our results are
not affected. These selections resulted in a final set of 11 indices.

\section{Results}

The results of the analysis are given in Table 1. Only the first two principal components are listed (the others each contribute $< 1\%$ of the total variance). Each column has
the eigenvector (principal component) with its associated eigenvalue (variance). Since the overall normalization is arbitrary, so is the sign of each PC, though the signs of
the individual elements in each PC are significant. While there are some common ``rules of thumb'' for deciding which PCs are significant (e.g., those with variance greater
than unity), there is no \emph{a priori} method for making such choices. To calculate these same PCs using different data, our normalizations (means and standard deviations)
must be used. For convenience, these values have also been included in the table.

PC1 is highly significant, accounting for $\sim 94\%$ of the total variance in the data. Most of the 11 individual components of PC1 have nearly equal weighting, though three of
them (the Balmer indices H$\beta$, H$\gamma_A$, and H$\delta_A$) have the opposite sign to the rest. This strongly suggests that PC1 is correlated with the overall GC metallicity.
This result is not surprising, since nearly all of the Lick indices primarily measure various metal absorption lines. We shall denote this metallicity as [m/H], since it is not
clear that either the iron abundance or the true ``total'' metallicity are being measured. We defer discussion of this issue until the next section.

PC2 accounts for just $\sim 3\%$ of the variance, and its normalized variance is much less than unity. Thus, it is not immediately clear whether it is significant. The primary
contributing index is H$\beta$, but since the other two Balmer lines do not strongly contribute, it is unlikely that PC2 represents a residual ``hot star'' component not
captured by PC1. It is possible that PC2 traces a nonlinear component of the relation between H$\beta$ and [m/H]. H$\beta$ is affected by intrinsic changes with [m/H] and
through \ion{Fe}{1} contamination of the index. The fact that nearly all the metal indices in PC2 contribute in the same sense as H$\beta$ is consistent with this idea, since
one would expect the magnitude of the correction to be greater for higher metallicities. If this interpretation of PC2 is correct, then it is of little physical interest. It
does suggest, however, that the ages of old ($> 1$ Gyr) stellar populations derived solely from H$\beta$ may be suspect. This nonlinear metallicity response of H$\beta$ could
cause spurious age spreads. Indeed, a wider spread in ages derived from H$\beta$ compared to those from H$\gamma$ and H$\delta$ has been observed (Puzia 2003) and has led to
the increasing use of higher-order Balmer lines in stellar populations work. These lines also have the advantage of being less susceptible to ``fill-in'' from background
emission.

In contrast to Rose \& Tripicco (1986), we find no evidence in our PCA for a second parameter for metal-rich GCs. However, their conclusions were based primarily on narrow,
medium resolution, non-Lick indices in the blue. None of these could be included in our analysis, so the present work does not provide evidence against such an effect. Indeed,
Figure 1 of Schiavon \etal~(2004) shows that the H$\delta_A$ values of the two metal-rich GCs known to have anomalously blue horizontal branches (BHBs), NGC 6388 and NGC 6441
(Rich \etal~1997), are $\ga 0.5$ \AA\ higher than other GCs of comparable metallicities. This indicates that a second parameter---in fact, the generic GC ``second parameter''
for HB morphology---is present in our sample, though it does not appear in our PCA. One possible explanation is the relative paucity of second parameter pairs of calibrating
GCs in the present study (J.~Rose 2004, private communication). The relatively wide Lick index bands may also disguise the effect of differing HB morphology on the Balmer
lines (see below).

Gregg (1994, hereafter G94) performed PCA on a set of nine indices in thirteen GCs, some of which measure individual absorption features (e.g., the G band) while others are
summed ``meta-indices'' of Fe or Balmer lines. The summed Fe index, named $\Sigma$Fe, is a sum of the equivalent widths of 23 Fe indices in the wavelength range 3550--6200 \AA.
We note that while G94 states that these are Fe indices, evidence from synthetic spectral calculations indicates that most low-resolution Lick indices are contaminated by
several elements and that many have poor sensitivity to the element for which they are named (Tripicco \& Bell 1995). G94 argues that the first three PCs are significant. His
first two PCs appear similar to those found in this study, with the former tracing metallicity and the latter dominated by the summed Balmer line strength $\Sigma$Balmer. He
suggests that his PC2 therefore represents a hot star component, a hypothesis which we rejected in our sample since neither H$\gamma$ nor H$\delta$ are prominent in our PC2.
Since G94 uses only $\Sigma$Balmer in his PCA, it is not possible to test whether his PC2 is driven solely by H$\beta$, as we find for our PC2. Additional evidence against his
PC2 measuring a hot star component is that NGC 6388 and NGC 6441 (see above) deviate in \emph{opposite} ways in a plot of metallicity vs. $\Sigma$Balmer (cf. Figure 3(d) in
G94). One caveat to this argument is that the G94 used somewhat narrower Balmer index passbands (12-18 \AA) than the Lick ones (which are $\sim 30$ \AA\ for H$\beta$, $\sim 20$
\AA\ for the higher-order ``F'' passbands, and $\sim 40$ \AA\ for the ``A'' passbands). However, we see no significant change in our PC2 in using the ``F'' or ``A'' definitions
(as noted above), suggesting this is less of a concern. Nevertheless, future measurements with narrower passbands may shed light on this discrepancy. Finally, G94 argues that
his PC3, despite having a normalized variance of only 0.1, is significant as well. PC3 is dominated by Ca, CN, and CH indices, and G94 suggests that it traces well-known
chemical variations of these elements and molecules within GCs (e.g., CN bimodality and Ca-Ti anticorrelation). The small number of sample clusters in G94 (13) makes it
difficult to test this suggestion further, though we see no evidence for scatter over observational errors in a PC1 vs.~CN$_2$ diagram.

\section{Discussion}

If, as proposed above, PC1 is related to [m/H], we would expect to see a correlation between PC1 and the metallicity of the GC. Figure 1 shows PC1 plotted against GC
metallicities from the Harris (1996, hereafter H96) catalog. H96 metallicities are the mean of all different published metallicity values for a given cluster, which come both
from spectroscopy and CMDs. H96 notes in the references to the catalog that the scale is essentially that of Zinn \& West (1984, hereafter ZW84), though with more modern
measurements it is becoming closer to that of Carretta \& Gratton (1997). This latter work used high-resolution spectra to establish a ``true'' [Fe/H] scale (it is not clear
that the ZW84 scale measures [Fe/H]), but unfortunately, the limited metallicity range of Carretta \& Gratton (1997) and the minimal overlap between our cluster samples makes
it impractical to directly compare our results to their metallicity scale. A similar problem exists with the recent work of Kraft \& Ivans (2003) who established a GC
metallicity scale using \ion{Fe}{2}. The motivation for this latter work is that \ion{Fe}{2} is expected to be less affected by possible non-LTE effects in modeling line
strengths than \ion{Fe}{1}, which was used, for example, by Carretta \& Gratton (1997).

Our use of [m/H] (rather than [Fe/H] or [Z/H]) to denote metallicity reflects our uncertainty as to the nature of the quantity measured by [m/H]. ZW84 GC metallicities were
generally estimated either directly from $Q_{39}$, a measure of the line blanketing from $\sim 3800-4000$ \AA, or indirectly from linear relations between $Q_{39}$ and several
strong absorption features (Ca K, G band, and Mg$b$). They calibrated the $Q_{39}$ scale against Cohen (1983 and references therein) metallicity estimates, which in turn were
primarily based upon high and low resolution measurements of Fe and Mg lines in GC red giants. Given that the primary contributions to $Q_{39}$ are made by Ca H+K, the CN B
band, and the general forest of Fe lines (Carretta \& Gratton 1997), we conclude that ZW84 metallicities are likely a nonlinear function of (at least) Fe, Mg, C, N, and Ca. The
abundances of these elements, and especially that of O, which dominates metallicity, are unknown for most GCs. Thomas, Maraston, \& Bender (2003) suggest that both their models
and the ZW84 scale estimate total metallicity since their model predictions for GC metallicities (based upon Fe and Mg lines) are consistent, within 0.1-0.2 dex, with the ZW84
ones. As noted above, ZW84 metallicities are a hybrid of many different elements, so both Thomas \etal~(2003) and ZW84 metallicities are unlikely to reflect actual [Z/H]
values. While the spectra of the old stellar systems (e.g., extragalactic GCs, galaxies) that are compared to these models are rarely of sufficient quality for the distinction
between [Fe/H], [Z/H], and [m/H] to make a practical difference in the deduced metallicities, it is an important conceptual point which could be significant if individual
chemical abundances are being considered.

Figure 1 suggests a strong, approximately linear relationship between PC1 and the H96 metallicities in the range $-1.8 \le$ [m/H] $\le 0$. A simple linear fit gives:
\begin{equation}
\textrm{[m/H]} = 0.157 \,\textrm{PC1} - 1.062
\end{equation}

The 1-$\sigma$ error in this fit is $\sim 0.13$ dex. However, the intrinsic scatter in PC1 metallicities may be smaller. As an example, in Figure 2 we have plotted PC1 against the
normalized Fe4383 index for our sample. It is clear that the correlation is much tighter than that between PC1 and H96 [m/H]. Excluding the single highest-metallicity point, the
1-$\sigma$ error is only $\sim 0.08$ dex (using equation 1 derived above). Thus, it appears that much of the observed scatter in the PC1-[m/H] relation may lie in the H96 [m/H]
values themselves, which likely have some spread due to the averaging of metallicity estimates using different methods with different systematic errors. This suggests that future
high-resolution work on Galactic GC metallicities, especially in the high metallicity regime, should allow us to refine the accuracy of the relation.

The evidence presented thus far indicates that PC1 may be used to estimate the metallicities of GCs from their integrated spectra. This use is in a similar spirit to that of
Brodie \& Huchra (1990). However, we have more indices (thus utilizing more of the available information) and the analysis can easily be extended to include additional indices.
Our relation also extends to higher metallicity, and much of the work on extragalactic GCs is focused on the metal-rich GCs. An eventual application to luminous early-type
galaxies with supersolar metallicities awaits further calibration beyond the current solar metallicity limit. In the analysis of old stellar populations, the use of PC1 is likely
to be preferable to the common procedure of estimating [m/H] from Lick index-index grids, as it uses many more metal indices and is largely \emph{independent} of SSP models. Note,
though, that some of the GC metallicities included in the H96 averages were calculated by isochrone fitting to CMDs and that these isochrones are based on SSP models.

We note that Proctor \etal~(2004) have separately developed an SSP-based method for estimating GC metallicities, ages, and [$\alpha$/Fe] abundances using a $\chi^2$-minimization
procedure with Lick indices and interpolated SSP model grids. We expect the two methods will offer complementary information in the study of extragalactic GCs, as the SSP method
can be utilized for young ages, supersolar metallicities, and different [$\alpha$/Fe] enhancement than found in GCs (with all of the accompanying model caveats).

As a check on these results, we calculated PC1 at two representative model metallicities ([Z/H] = $-1.35, -0.33$; [$\alpha$/Fe] = +0.3) in the SSP models of Thomas \etal~(2003).  
The resulting PC1 values were -1.72 and 4.78; using (1) above, these correspond to [m/H] values of -1.33 and -0.31, which agree astonishingly well ($\sim 0.02$ dex) with the input
metallicities. This agreement is all the more surprising when one considers that the SSP metallicity scale and that of the GCs might not be expected \emph{a priori} to be the same
(see above for further discussion).

Several caveats apply to the accuracy and use of the relation. First, outside of the listed ranges of validity, the linear fit should be used with caution. Indeed, the paucity
of points in the range $-0.5 \le$ [m/H] $\le 0$ should be considered; the accuracy of [m/H] estimates in this range may be somewhat lower than stated for the relation as a
whole. Second, since all of the GCs in our sample have very old ages ($\ga 11$ Gyr), it is unclear how well PC1 would estimate metallicities for young or intermediate-age GCs.
Since the majority of the Lick indices were defined to measure metal absorption lines in the spectra of G dwarfs and K giants (which dominate the integrated spectrum of an old
stellar population) PC1 is unlikely to be useful for ages $\la 1$ Gyr. Encouragingly, PC1 appears to give metallicities consistent with those derived from Thomas \etal~(2003)
SSPs for several intermediate-age GCs in the LINER E4 NGC 1052 (Pierce \etal~2004). However, these same SSP models give supersolar metallicities, [Z/H] $\sim +0.5$, for two
$\sim 2$ Gyr GCs in the merger remnant elliptical galaxy NGC 3610 (Strader, Brodie, \& Forbes 2004), while the PC1-derived metallicities are much lower at $\sim -0.3$ dex.  
The extremely high metallicities derived from these SSPs are somewhat surprising, but it is not clear whether the discrepancy results primarily from the models or the lack of
calibrating metal-rich, intermediate-age GCs in our PCA. Calibrating CMDs and integrated spectra for the Magellanic Cloud GCs would allow a good local test of this issue,
though only at subsolar metallicities. Third, stellar populations with different abundance patterns than Galactic GCs (which typically have [$\alpha$/Fe] $\sim +0.3$) might be
expected to follow a different relation. Fourth, caution should also be used in estimating metallicities for composite stellar populations (e.g., galaxies). The resulting
metallicities would be luminosity-weighted averages of the contributions of individual subpopulations (this is also true for SSP model comparisons). Finally, this relation
should strictly only be used if all the included indices have been measured. However, for a subset of these indices, or others in the wavelength region $\sim 4000-5500$ \AA,
the authors can furnish an applicable metallicity calibration upon request.

As a side note, G94 also calibrated his $\Sigma$Fe parameter for use as a metallicity indicator, though the relation was fitted with only six calibrating GCs. It appears linear
over a wide range of metallicities, but has the practical disadvantage that as currently defined, $\Sigma$Fe requires the measurement of indices over a very large wavelength range
(3550--6200 \AA).

The PCA presented here could easily be extended to include the remainder of the Lick indices and features not in the Lick system (e.g., the 4000-\AA\ break) in the spirit of G94.
It could even be applied to the entire spectrum, as has become common in large galaxy surveys (Madgwick \etal~2003). As it stands, our analysis offers a potentially powerful tool
for estimating the metallicities of extragalactic GCs.

The present study indicates that the Lick indices of Galactic GCs are (at least) a two-parameter family, though only the first parameter (metallicity) appears significant in
our PCA. The second parameter (HB morphology) could emerge more clearly in future studies if narrow indices are used or a larger sample of calibrating clusters is available,
especially ones covering a range of HB morphology at a given metallicity. At some level other parameters (e.g., abundance variations along the red giant branch) must exist,
but the key issue is at what level of detail they become apparent in the integrated spectrum of a GC. For example, Schiavon \etal~(2004) find that BHBs can be distinguished
from younger ages in the spectra of Galactic GCs through the use of SSPs and multiple Balmer lines, provided the spectra are of sufficiently high S/N ($\sim 30$/\AA). It is
difficult to achieve such high S/N even for the brightest old GCs beyond the Local Group (except for the closest galaxies) in an entire night on an 8--10-m class telescope.
While BHBs may have a pronounced effect on the Balmer lines of metal-rich GCs that is detectable at a lower S/N, it may not be possible to differentiate this from younger
ages. If high S/N or moderate resolution spectra are required to detect such effects in the integrated spectra of extragalactic GCs, then these second or higher-order
parameters may be of limited use in studying stellar populations.

In addition, the quality and size of the calibrating data set (Schiavon \etal~2004), as well as the sensitivity of PCA, allows us to draw firmer conclusions than previously
possible about differences in abundance patterns among Galactic GC subpopulations, e.g., between the metal-poor and metal-rich clusters. Our determination that such differences
are minimal represents a constraint on the formation of these subpopulations. They must have been formed through very similar physical processes, or at least ones which left
similar chemical imprints.

\acknowledgements

We thank the referee Jim Rose for insightful comments, Ricardo Schiavon, Jim Rose, and St\'{e}phane Courteau for use of their indices in advance of publication, and 
Mike Beasley for helpful discussion. We acknowledge support by the National Science Foundation through Grant AST-0206139 and a Graduate Research Fellowship (JS).

\newpage

\epsfxsize=14cm
\epsfbox{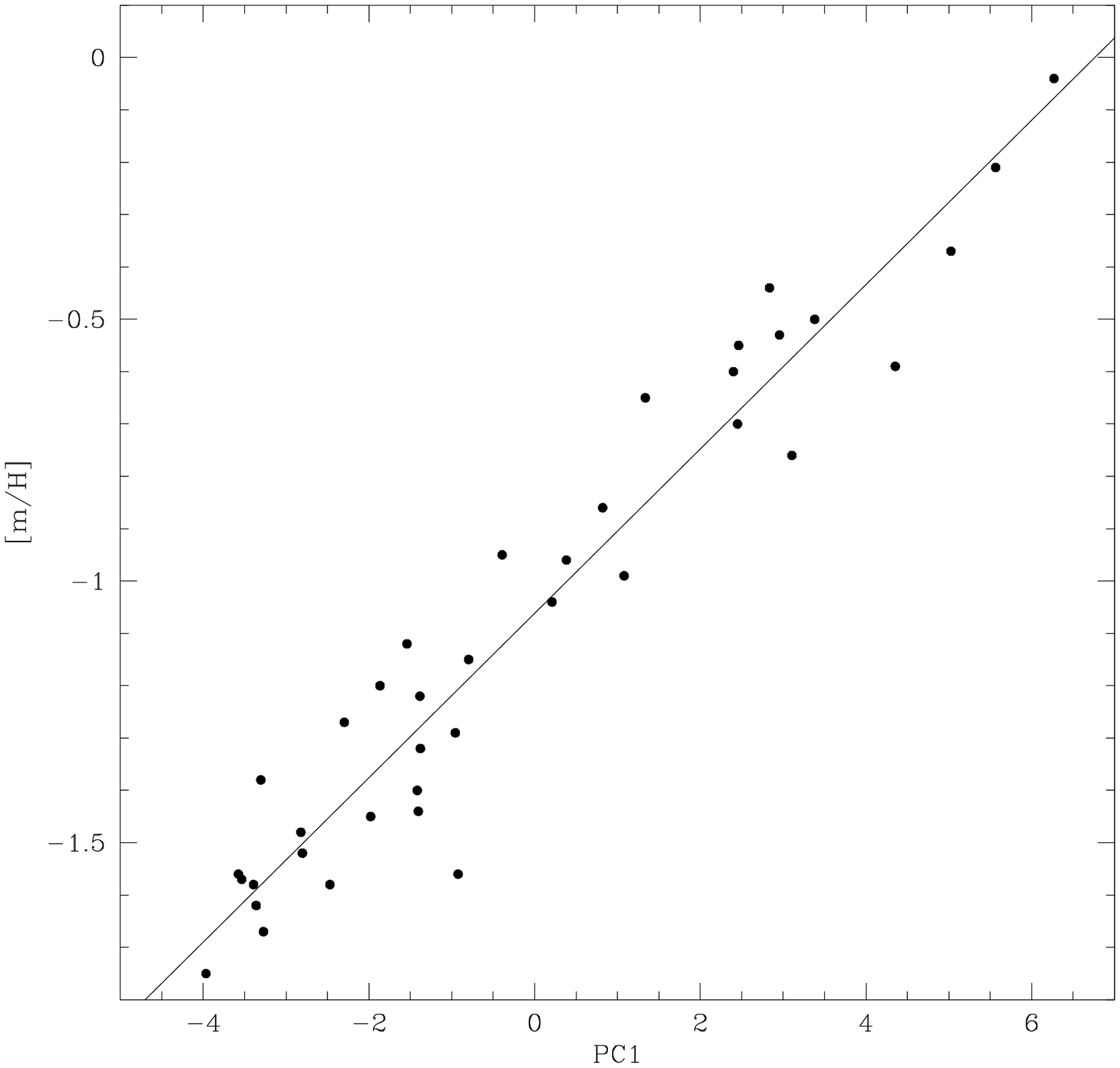}
\figcaption[strader.fig1.ps]{\label{fig:fig1} GC metallicities from Harris (1996) vs. PC1. A linear fit (see text) is overplotted. The $1-\sigma$ error in the fit is 0.13 dex.}

\newpage

\epsfxsize=14cm
\epsfbox{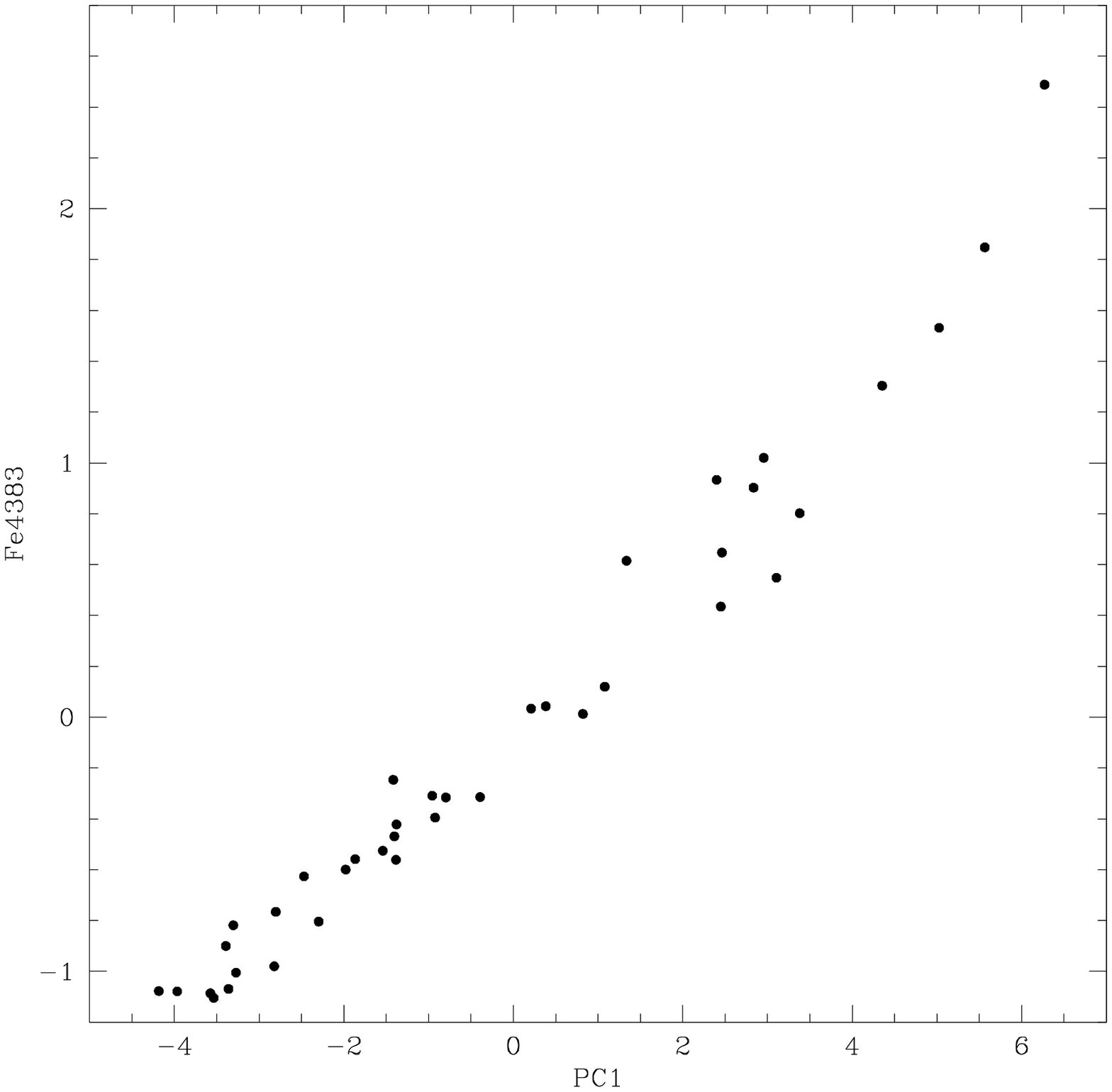}
\figcaption[strader.fig2.ps]{\label{fig:fig2} Normalized Fe4383 vs.~PC1. The scatter in this diagram is smaller than that in Figure~1, suggesting
much of the error in the linear fit is intrinsic scatter in the Harris (1996) metallicities.}

\newpage

\begin{deluxetable}{lrrrl}
\tablewidth{0pt}
%\rotate
\tablecaption{Results of PCA
        \label{tab:pcares}}
\tablehead{Label/Index & PC1 & PC2 & mean & sigma}
\startdata

Variance	&	8.27	&	0.27	&		&		\\
$\%$ of Total	&	0.94	&	0.03	&		&		\\
\hline
H$\delta_A$	&	-0.247	&	 0.086 &	1.436	&	2.184	\\
CN$_2$		&	 0.302  &	 0.203 &	-0.004	&	0.054	\\
Ca4227		&	 0.297	&	-0.057 &	0.399	&	0.193	\\
G4300		&	 0.270	&	-0.348 &	3.042	&	1.302	\\
H$\gamma_A$	&	-0.281	&	 0.208 &	-1.353	&	2.550	\\
Fe4383		&	 0.306	&	 0.154 &	1.844	&	1.150	\\			
H$\beta$	&	-0.294	&	 0.740 &	2.017	&	0.389	\\
Mg$b$		&	 0.321	&	 0.196 &	1.722	&	0.889	\\
Mg$_2$		&	 0.332	&	 0.203 &	0.096	&	0.053	\\
Fe5270		&	 0.327	&	 0.242 &	1.409	&	0.502	\\
Fe5335		&	 0.329	&	 0.271 &	1.134	&	0.436	\\

\enddata
\end{deluxetable}

\end{document}